# Self-Financing, Replicating Hedging Strategies
*an incomplete analogy with thermodynamics*

Joseph L. McCauley
Physics Deptartment
University of Houston
Houston, Texas 77204
jmccauley@uh.edu

## Abstract

In the theory of riskfree hedges in continuous time finance, one can start with the delta-hedge and derive the option pricing equation, or one can start with the replicating, self-financing hedging strategy and derive both the delta-hedge and the option pricing partial differential equation. Approximately reversible trading is implicitly assumed in both cases. The option pricing equation is not restricted to the standard Black-Scholes equation when nontrivial volatility is assumed, but produces option pricing in agreement with the empirical distribution for the right choice of volatility in a stochastic description of fluctuations. The replicating, self-financing hedging strategy provides us with an incomplete analogy with thermodynamics where liquidity plays the role of the heat bath, the absence of arbitrage is analgous to thermal equilibrium, but there is no role played by the entropy of the returns distribution, which cannot reach a maximum/equilibrium. We emphasize strongly that the no-arbitrage assumption is *not* an equilibrium assumption, as is taught in economics, but provides only an incomplete, very limited analogy with the idea of thermal equilibrium.

We define an approximately reversible trade as one where you can reverse your buy or sell order over a very short time interval (on the order of a few seconds or ticks) with only very small percentage losses, in analogy with approximately reversible processes in laboratory experiments in thermodynamics. All that follows assumes that approximately reversible trading is possible, although reversible trading is the exception in the market.

Consider a dynamic hedging strategy $(\phi,\psi)$ defined as follows [1]. Suppose you're short a call at price C(p,t). To cover your bet that the underlying stock price will drop, you simultaneously buy $\phi$ shares of the stock at price p by borrowing $\psi$ dollars from the broker (the stock is bought on margin, e.g.). In general, the strategy consists of holding $\phi(p,t)$ shares of stock at price p, a risky asset, and $\psi(p,t)$ shares of a money market fund at initial price m=1 Euro/share, a riskless asset (with fixed interest rate r) at all times t≤T during the bet, where T is the strike time. At the initial time $t_o$ the call is worth

$$C(p_o,t_o) = \phi_o p_o + \psi_o m_o$$

(1)

where $m_o$=1 Euro. Assuming that $(\phi,\psi)$ are twice differentiable functions, the portfolio is self-financing if, during dt,

$$d\phi p + d\psi m = 0$$

(2)

so that

$$dC = \phi dp + \psi dm$$

(3)

where dm=rmdt. In (3), dC and dp and stochastic variables because p(t+dt) and C(t+dt) are unknown and random at time t when p(t) and C(p,t) are observed. Viewed as C(p,m) eqn. (3) tells us that

$$\phi = \frac{\partial C}{\partial p}$$

(4)

but we will return to this below. Next, we want the portfolio in addition to be 'replicating', meaning that the twice differentiable functional relationship

$$C(p,t) = \phi(p,t)p + \psi(p,t)m$$

(5)

holds for all later (p,t) up to the strike, and p is the known price at time t (for a stock purchase, we can take p to be the ask price). Equation (5) expresses the idea that holding the stock plus money market in the combination ($\phi$,$\psi$) is equivalent to holding the call. The strategy ($\phi$,$\psi$), if it can be constructed, defines a 'synthetic call': the call at price C is synthesized by holding a certain number $\phi$>0 shares of stock and $\psi$<0 of money market at each instant t and price p(t). These conditions, combined with Ito's lemma, predict the option pricing equation and therefore the price C of the call [1]. An analogous argument can be made to construct synthetic puts, where covering the bet made by selling the put means shorting $\phi$ shares of the stock and holding $\psi$ dollars in the money market. In this paper we do not criticize the assumption of continuous time finance, with ticks on the order of seconds, but leave that for a later paper.

Several assumptions are necessary in order for the synthesis to work. We state here only the assumptions central for the analogy with thermodynamics. One is that transaction costs are negligible. Another is that the 'liquidity bath' is large enough that borrowing the money, selling the call and buying the stock are possible approximately instantansously (during a few ticks in the market) without affecting the price of either the stock or call, or the interest rate r. That is, the desired margin purchase is assumed to be possible approximately reversibly in real time through your discount broker on your Ibook or PC. This will not be possible if the number of shares involved is too large, or if the market crashes or balks (because limit bid/ask prices have too large a spread) at selling ϕ shares approximately reversibly (the 'liquidity bath' dries up in either case).

Starting with the sde for the stock price

$$dp = Rpdt + \sigma(p,t)pdB$$

(6)

where B(t) defines a Wiener process, with <dB>=0 and <dB²>=dt, and using Ito's lemma [1,3] we obtain the sde

$$dC = (\dot{C} + \sigma^2 p^2 C'' / 2)dt + C'dp$$

(7)

On the other hand, from (3) and Π=−ψm, we have

$$dC = \phi dp - d\Pi = \phi dp - r\Pi dt = \phi dp - r(-C + \phi p)dt$$

(8)

As is pointed out in [4], r need not be the riskfree interest rate and, in practice, is not but usually is a few percentage points higher. Equating coefficients of dp and dt in (7) and (8) we obtain

$$\phi = \frac{\partial C}{\partial p}$$

(9)

which agrees with having treated ϕ as differentiable above, and the option pricing pde

$$\frac{\partial C}{\partial t} + \frac{\sigma^2(p,t)p^2}{2}\frac{\partial^2 C}{\partial p^2} = -rp\frac{\partial C}{\partial p} + rC$$

(10)

With (p,t)-dependent volatility $\sigma^2(p,t)$, the pde (10) is not restricted to Black-Scholes/Gaussian returns, but includes the exponential model of returns [4] as well, although to obtain the exponential model we must start with the sde

$$dp = (R + \sigma^2 / 2)dt + \sigma(p,t)pdB(t)$$

(11)

The reason for this difference with standard theory is empirics: we cannot determine 'microscopically' what is the expected return R, the coefficient of dt in (6), but must consult instead the empirical distribution of returns [4]. The empirical distribution of returns, which is exponential, is consistent with a p-independent expected return R appearing in the Fokker-Planck equation [4],

$$\frac{\partial P(x,t)}{\partial t} = -\frac{\partial (R(t)P(x,t))}{\partial x} + \frac{1}{2}\frac{\partial^2 (D(x,t)P(x,t))}{\partial x^2}$$

(12)

where $x=\ln(p(t))/p(0))$ is the empirically-demanded logarithmic return [4], with corresponding stochastic integral equation [3,4]

$$\Delta x = \int_{t}^{t+\Delta t} R(s)ds + \int_{t}^{t+\Delta t} \sqrt{D(x(s),s)}dB(s)$$

(13)

and where the stochastic integral on the right in (13) is path-dependent but can be solved by iteration, just as in the theory of ordinary differential equations whenever a Lipshitz condition is satisfied [3].

Given the pde (10), pricing the call is then as follows: one simply solves the option pricing pde but first rewritten via a transformation in terms of returns x, as is shown in [4], and one then applies the forward-time initial condition C(p,T) =max(p-K) at strike time T (K is the strike price), using $p=p_o e^x$.

Equations (3), (4) and (5) define a Legendre transform. We can now make an incomplete analogy with thermodynamics, even though the market distribution is not in equilibrium, and even though our variables are stochastic ones. Eqn. (2) is like a generalized Gibbs-Duhem relation and Π=-ψm is analogous to a thermodynamic potential. If we ask for a thermodynamic potential where extensive variables appear as differentials (we assume that (ϕ,ψ) are analogous to extensive variables in the number of shares held) then eqn. (5) is analogous to a thermodynamic potential Φ =TS$_o$=E+PV-μN with entropy S$_o$ held constant (reversible adiabatic process), so that dΦ=VdP-Ndμ is analogous to (3). So far, this is just empty formalism. The essential part is that the assumption of adequate liquidity is analogous to the heat bath, and absence of arbitrage possibilities is analogous (but certainly not equivalent) to thermal equilibrium, where there are no correlations: one can't get something for nothing out of the heat bath because of the second law. In the financial case, arbitrage is imposible systematically in the absence of correlations.

If we next consider fluctuations about thermal equilibrium, then the analog of a riskfree portfolio (Π=-C+C'p, regarded as a portfolio of a call and the underlying stock, is riskfree [2,4]) is the vanishing of the mean square fluctuation of the thermodynamic potential Φ in the limit of an infinite system far from a phase transition (however, this thermal analogy breaks down because fluctuations in the corresponding analogues ϕp and C of thermodynamic potentials do *not* vanish). We neither see nor suggest any analogy with phase transitions. The better analogy is that large trades violate the heat bath/liquidity assumption, just as taking too much energy out of the system's environment violates the assumption that the heat bath remains approximately in equilibrium in thermodynamics.

In contrast with the *qualitative* difference between the thermodynamic potential $\Phi=TS_o$ and the call price C, we can write down and discuss the (true) Gibb's entropy S(t) of the empirical returns distribution P(x,t). The possibility of arbitrage would correspond to a lower entropy (reflecting correlations in the empirical returns distribution), as was discussed by Zhang [5]. One severe weakness of the above ‚thermodynamic analogy' is that the entropy S(t) of the returns distribution P(x,t) does not appear in (5), and whereas $S_o$ is constant the *real* entropy of the market,

$$S(t) = -\int_{-\infty}^{\infty} P(x,t)\ln P(x,t)dx$$

(14)

is always increasing because P(x,t) is diffusive. Note that S(t) can never reach a maximum due to the time-dependent diffusion coefficient D(x,t) and the lack of finite upper and lower bounds on x. Even with a t-independent volatility D(x) and expected stock rate of return R, the solution of (12) describing statistical equilibrium,

$$P(x) = \frac{C}{D(x)} e^{2\int \frac{R(x)}{D(x)} dx}$$

(15)

would be possible after a long enough time only if a ‚particle' is confined, either by a potential U(x,t) or by other constraints, e.g. by a constraint where the ‚particle' bounces back and forth between two walls $p_{min} \leq p \leq p_{max}$. This is not what is taught in standard economics texts.

Arbitrage possibilities would correspond to correlations (history-dependence in P(x,t)), whereas the no-arbitrage condition (which is essentially the ‚efficient market hypothesis', or EMH) is satisfied by statistical independence in P(x,t). Our model of volatility of empirical returns [4] is based on the assumption of statistical independence. The presence of correlations in a returns distribution would imply a lower entropy than in the case of no correlations, but entropy has been ignored in the economics literature where the emphasis has been (mis-)placed on the extreme (and empirically incorrect) notions of perfect foresight, instant information transfer [6] and utility maximization [7], instead of *loss of information.*

We cannot stress too much that the 'no-arbitrage' condition is *not*, as is claimed in papers and books on economics and finance [2], an equilibrium condition, dynamically seen. In Fama's famous paper on the EMH [6] he assumes that Martingale expectation values of random variables represent 'market equilibrium' even though those expectation values are time-dependent, and even though the market is generally far from dynamic equilibrium: total excess demand never vanishes due to outstanding limit orders. That limit orders cannot be filled is equivalent to saying that agents are dissatisfied with bid/ask prices that are offered, and so prices fluctuate approximately statistically-independently, never reaching any equilibrium. Again, no-arbitrage in very liquid markets is not an equilibrium condition but provides at best only a very weak and incomplete *analogy* with thermal equilibrium, *where in the latter nothing changes with time*.

The assumption of 'no market impact' (meaning enough liquidity) during trading is an approximation that is limited to very small trades in a heavily-traded market (and is violated when, e.g., Deutsche Bank takes a very large position in Mexican Pesos or Czec Crowns; see Dacorogna et al [8] for a discussion of the effects of large limit orders). In thermodynamics, the reservoir's temperature can be treated as unaffected by energy exchanges with the system only to 0th order, where the first order entropy change is zero ($\delta S=0$ defines a reversible process). However, the total entropy change of system plus reservoir is given by [9]

$$\Delta S = \delta S + \delta^2 S + ... + \delta^n S + ...$$
(16)

where the second order change is positive, $\delta^2 S>0$, guaranteeing stability of thermal equilibrium (stability fails at a phase transition, e.g.), because the temperature difference between the reservoir and the system is finite, otherwise no energy exchange, no matter how small, can occur approximately reversibly.

No comparable description or correction to first order effects exists in option pricing. However, reversible trading *is* consistent with the observed empirical fat-tailed distribution (and therefore with large fluctuations) for a choice of volatility $\sigma^2(p,t)$ that permits description of the empirical distribution by a Fokker-Planck equation. We have shown in [4], for the empirical distribution P(x,t) for interday trading of bonds and foreign exchange, written in terms of returns x=ln(p(t)/p(0)), that the volatility is given by $\sigma^2(p,t) = D(x,t)=bv(x-\delta)$ where b is constant but v and δ are time-dependent.

The idea of synthetic options, based on equation (5), led to so-called 'portfolio insurance', which was based implicitly on the assumption of approximately reversible trading, that agents would always be there to take the other side of a desired trade at approximately the price wanted. In October, 1987, the market crashed, meaning that liquidity dried up, and many people who had believed (without really thinking about the implicit assumption of liquidity [10]) that they were ‚insured' lost money. The idea of portfolio insurance was based on an excessive belief in the mathematics of approximately reversible trading combined with the expectation that the market will go up, on the average (R>0), but ignoring the (unknown) time scales over which downturns and recoveries occur, if ever. The strategy of a self-financing, replicating hedge requires an agent to buy on the way up and sell on the way down. This strategy led to buying high and selling low, which is not usually a prescription for success.

Another example of misplaced trust in neo-classical economic beliefs is the case of LTCM, where it was assumed that prices would always return to historic averages, in spite of the lack of ‚springs' in the market (there is no restoring force in (6), no evidence for a stabilizing ‚invisible hand'), and correspondingly throwing good money after bad until the Gamblers' Ruin [11] ended the game. One does not know yet the real story of which assumptions led to the recent disaster of Enron, other than (as with LTCM) the belief that unregulated free markets are stable, but apparently they are not. The entropy (14) is always increasing, never reaching a maximum, and is consistent with very large fluctuations that have unknown and completely unpredictable relaxation times. The neo-classical economists have it all wrong: one cannot have *both* deregulation *and* stable equilibrium. One might *at best* have *either* stability *or* total lack of regulations (the individual moral contraint in agents that Adam Smith [12] presummed necessary for stabilizing the so-called ‚invisible hand' is completely lacking in modern global capitalism, if ever it was present). Based on the observed instability of unregulated free markets as reflected by liquid market statistics, we predict that more deregulation will lead to more financial disasters. Globalization since the downfall of the USSR is a completely uncontrolled global experiment whose outcome cannot be correctly predicted on the basis of either past history (statistics) or neo-classical economic theory, but with LTCM, Enron and Argentina as examples of some of the recent consequences of deregulation one should not be optimistic.

**Acknowledgement**

I'm particularly grateful to H. Joseph Hrgovcic for suggesting to me (in 10/2000) that there may be an analogy between option pricing and Caratheorody's formulation of thermodynamics (which I have not investigated here). I'm grateful to Doyne Farmer for very friendly email correspondence, for pointing out to me that pitfalls in the assumption of reversible trading were discussed in an early version of his paper [13], and also for telling me earlier (at a very pleasant small complexity meeting at UNAM in Mexico City in 10/01) of his interest in thermodynamic analogies in economics. Awareness of these expectations led me to notice the entropy-free thermodynamic analogy while working through and trying to understand the idea of the replicating, self-financing hedge. I'm also grateful to Gemunu Gunaratne for email discussions and encouragement while I am living in and writing from an Austrian mountain village with a beautiful view.

## References

1. Baxter, M. and Rennie,A. 1996 Financial Calculus (Cambridge, Cambridge)

2. Bodie, Z. and Merton, R.C., 1998 Finance (Saddle River, Prentice-Hall)

3. Arnold, L. 1992 Stochastic Differential Equations (Krieger, Malabar, Fla.)

4. McCauley, J.L. and Gunaratne, G.H. 2002 An Empirical Model of Volatility of Returns and Option Pricing (submitted)

5. Zhang, Y.-C. 1999 Physica A269, 30

6. Fama, E.F. 1970 J. Finance 383

7. Varian, H.P, 1992 Microeconomic Analysis (Norton, NY)
8. Dacorogna, et al M. 2001 An Intro. To High Frequency Finance (New York, Academic)

9. Callen, H.B. 1985 Thermodynamics (Wiley, NY)

10. Jacobs, B.I. 1999 Capital Ideas and Market Realities : Option Replication, Investor Behavior, and Stock Market Crashes (Blackwell, )

11.  Billingsley, P. 1983 American Scientist 71, 392

12. Smith, A. 2000 Wealth of Nations (Princeton, Princeton)

13. Farmer, J.D. 1994 Market Force, Ecology, and   Evolution (preprint, the original version)